\documentclass[prd,nofootinbib,amsfonts,notitlepage]{revtex4-1}
\usepackage[utf8]{inputenc}
\usepackage[T1]{fontenc}
\usepackage{hyperref}
\usepackage{graphicx,bm,amsmath,color}
\usepackage{bbold}
\usepackage{wasysym}
\usepackage{verbatim}
\usepackage{amssymb}
\usepackage{epstopdf}
\usepackage{animate}
\usepackage{xmpmulti}
\usepackage{centernot}
\usepackage{multirow}
\usepackage{tikz}
\usepackage[normalem]{ulem} 
\makeatletter

\newcommand{\be}{\begin{equation}}
\newcommand{\ee}{\end{equation}}
\newcommand{\beq}{\begin{eqnarray}}
\newcommand{\eeq}{\end{eqnarray}}
\newcommand{\ba}{\begin{align}}
\newcommand{\ea}{\end{align}}

\begin{document}

\title{Relativistic deformed kinematics from momentum space geometry}
\author{J.M. Carmona}
\affiliation{Departamento de F\'{\i}sica Te\'orica and Centro de Astropartículas y Física de Altas Energías (CAPA),
Universidad de Zaragoza, Zaragoza 50009, Spain}
\author{J.L. Cort\'es}
\affiliation{Departamento de F\'{\i}sica Te\'orica and Centro de Astropartículas y Física de Altas Energías (CAPA),
Universidad de Zaragoza, Zaragoza 50009, Spain}
\author{J.J. Relancio}
\email{jcarmona@unizar.es, cortes@unizar.es, relancio@unizar.es}
\affiliation{Departamento de F\'{\i}sica Te\'orica and Centro de Astropartículas y Física de Altas Energías (CAPA),
Universidad de Zaragoza, Zaragoza 50009, Spain}

\begin{abstract}
We present a way to derive a deformation of special relativistic kinematics (possible low energy signal of a quantum theory of gravity) from the geometry of a maximally symmetric curved momentum space. The deformed kinematics is fixed (up to change of coordinates in the momentum variables) by the algebra of isometries of the metric in momentum space. In particular, the well-known example of $\kappa$-Poincar\'e kinematics is obtained when one considers an isotropic metric in de Sitter momentum space such that translations are a subgroup of the isometry group, and for a Lorentz covariant algebra one gets the also well-known case of Snyder kinematics. We prove that our construction gives generically a relativistic kinematics and explain how it relates to previous attempts of connecting a deformed kinematics with a geometry in momentum space.
\end{abstract}

\maketitle

\section{Introduction}
There have been many attempts to avoid the problems of inconsistency between general relativity and quantum field theory, including string theory~\cite{Mukhi:2011zz,Aharony:1999ks,Dienes:1996du}, loop quantum gravity~\cite{Sahlmann:2010zf,Dupuis:2012yw}, supergravity~\cite{VanNieuwenhuizen:1981ae,Taylor:1983su}, or causal set \mbox{theory~\cite{Wallden:2010sh,Wallden:2013kka,Henson:2006kf}}. Unfortunately, it is difficult to extract from these first-principle theories the experimental predictions that could tell us which is the correct approach to a quantum gravity theory (QGT). In this sense, doubly special relativity (DSR, see Ref.~\cite{AmelinoCamelia:2008qg} for a review) was proposed as a low-energy limit of a QGT that could be testable by experimental observations. 

DSR is an example of a \emph{deformed} kinematics, by which we mean any modification of the kinematics of special relativity which depends on a high-energy scale $\Lambda$, in such a way that special relativity (SR) is recovered in the limit $\Lambda\to \infty$. In the particular case of DSR, this high-energy scale is identified with the Planck energy, and the symmetries of SR are deformed in such a way that the Planck energy, as well as the speed of light, is an invariant quantity in any reference frame. This is done by means of a nonlinear implementation of Lorentz transformations which now depend on both quantities. The new transformations relate inertial observers describing a deformed kinematics subject to a relativity principle, i.e., a \emph{relativistic deformed kinematics} (RDK). This kinematics contains, generically, two non-independent ingredients: a deformed dispersion relation (DDR), and a conservation law which involves a deformed composition law (DCL) for momenta. The need to change the standard addition of momenta to a nonlinear composition stems from the necessary compatibility with the relativity principle which is implemented by the nonlinear Lorentz transformations, and it is related to a modification of the Poincaré translations, leading to nonlocal effects known as relative locality~\cite{AmelinoCamelia:2011bm}.

As we will show in this work, a DDR and a DCL can also be introduced through a geometry in momentum space.
The idea of considering a curved momentum space was firstly proposed by Born in the 1930's~\cite{Born:1938}, and also discussed some years later by Snyder~\cite{Snyder:1946qz}, as a way to avoid the ultraviolet divergences in quantum field theory. More recently, a curved momentum space was introduced as dual of a noncommutative spacetime in connection with quantum-deformed Hopf algebras describing space-time symmetries~\cite{Majid:1999tc}. In the particular case of the $\kappa$-Poincaré Hopf algebra~\cite{Lukierski:1991pn}, the associated $\kappa$-Minkowski noncommutative spacetime~\cite{Majid1994} allows one to deduce a de Sitter geometry for momentum space~\cite{KowalskiGlikman:2002ft}.

The mathematical language of Hopf algebras~\cite{Majid:1995qg} represents a way to go beyond the SR kinematics, where the DDR is given by the Casimir of the deformed Poincaré algebra in a certain basis, and the DCL is given by a nontrivial coproduct operation (see, for example, Ref.~\cite{KowalskiGlikman:2002we} and Sect. IV of Ref.~\cite{Carmona2016}). This \emph{algebraic} approach relates a deformed kinematics with a geometry in momentum space that is obtained from the Hopf algebra structure. In this paper we are going to consider instead a \emph{geometric} approach, for which the deformed kinematics is related to fundamental geometric objects, such as the metric in momentum space, the connection, or (as we will see, more fundamental), the tetrad.

The geometric approach has also been investigated previously~\cite{AmelinoCamelia:2011bm,AmelinoCamelia:2011pe,Lobo:2016blj}.
In  Ref.~\cite{AmelinoCamelia:2011bm}, a proposal for a relation between momentum space geometry and a deformed kinematics was given. In this proposal, the dispersion relation is defined as the square of the distance from the origin to a point $p$, and the DCL is related to a non-metrical connection. Unfortunately, this work does not mention Lorentz transformations, and indeed it seems hard to implement a relativity principle in a systematic way. The authors of Ref.~\cite{Lobo:2016blj}, however, considered a possible correspondence between a deformed composition law and the isometries of the momentum metric corresponding to translations (in the sense that they do not leave invariant the origin). The homogeneous isometries were then identified as the Lorentz transformations, and in such a way, one may guess that a relativity principle might hold: the DDR (defined by a metric) would be compatible with the DCL (which is associated to an isometry of the same metric), and the composition of momenta would be compatible with Lorentz transformations. This is what the authors of the previous work assure, indicating that ``there is a well-defined commutation rule between both operations (since they generate a group).'' As one would want 10 isometries (4 translations and 6 Lorentz boosts), one has to restrict the possible momentum geometry to a maximally symmetric space. Then, there are only 3 possibilities: Minkowski, de Sitter or anti-de Sitter space. 

However, Ref.~\cite{Lobo:2016blj} did not formulate a clear way to obtain the composition law, because there are in fact many isometries that do not leave invariant the origin; an additional ingredient is needed to characterize them. Also, the argument by which one gets a deformed kinematics which is compatible with the relativistic principle is not really clear, since, as we will see, it is necessary to discuss the implementation of Lorentz transformations for a set of two momenta.

Our purpose in this paper will be to clarify the geometric approach by making concrete this proposal. In particular, we will present a precise way to understand the deformed composition law: it can be associated to translations, but in such a way that their generators form a specific algebra inside the algebra of isometries of the momentum metric. Moreover, we will give a detailed argument showing how the construction we propose gives a relativistic deformed kinematics. This will be the subject of Sec.~\ref{sec:derivation}.

It is obvious that this new approach gives the usual kinematics in the SR case: since the SR momentum space is flat, the square of the distance from the origin to a point $p$ is a quadratic relation, and the conservation law and Lorentz transformations, understood as the isometries of the Minkowski metric, are both linear. In Sec.~\ref{sec:examples}, we will see that the very much studied $\kappa$-Poincaré model is obtained for a de Sitter momentum space, when the generators of translations satisfy an isotropic closed algebra ($\kappa$-Minkowski), while the DCL of Snyder~\cite{Battisti:2010sr}
is obtained for the same geometry, when the Poisson bracket of two translation generators is proportional to the Lorentz generator. We will also see that the generalization of these two kinematics, the so-called hybrid models~\cite{Meljanac:2009ej}, can also be obtained within this approach.
This will allow us to compare the geometric and algebraic approaches to a deformed relativistic kinematics.

In order to complete the discussion, in Sec.~\ref{relative_locality} we will show the correspondence between our prescription and the relative locality framework posed in Ref.~\cite{AmelinoCamelia:2011bm}. Finally, in Sec.~\ref{conclusions} we will present our conclusions and future prospects.

\section{Derivation of a deformed kinematics from the geometry of a maximally symmetric momentum space}
\label{sec:derivation}

A deformed kinematics is a combination of a deformed (with respect to special relativity) relation between the energy and the momentum of a free particle (deformed dispersion relation) and a deformation of the expression of the total energy-momentum of a system of free particles in terms of the energy-momentum of each of the particles (deformed composition law). The combination of a DDR and a DCL defines a RDK if it is possible to find a (nonlinear) implementation of Lorentz transformations such that if one has a set of particle momenta in an initial and a final states satisfying the dispersion relation and the conservation of total energy-momentum, the set of Lorentz transformed particle momenta will satisfy the same dispersion relations and energy-momentum conservation law (relativity principle). This is just a generalization of the Poincaré invariance of the special relativistic kinematics with a ten-parameter group of transformations relating different inertial frames.

\subsection{Definition of the deformed kinematics}

The geometry of a maximally symmetric four-dimensional momentum space will be defined by a ten-parameter group of transformations which leave the metric invariant (isometries). This leads to explore the possibility to derive a RDK from the geometry of a maximally symmetric momentum space.

We start by considering a metric $g_{\mu\nu}(k)$ in momentum space\footnote{The simplest form for the metric is $g_{\mu\nu}(k)=\eta_{\mu\nu} \pm k_\mu k_\nu/\Lambda^2$, where $\Lambda$ is the energy scale of deformation and the two signs correspond to de Sitter and anti-de Sitter.} and isometries $k\to k'$ defined by the system of equations
\be
  g_{\mu\nu}(k') \,=\, \frac{\partial k'_\mu}{\partial k_\rho} \frac{\partial k'_\nu}{\partial k_\sigma} g_{\rho\sigma}(k) .
\ee
We take a system of coordinates such that $g_{\mu\nu}(0)=\eta_{\mu\nu}$, where $\eta_{\mu\nu}$ is the Minkowski metric and we write the ten-dimensional group of isometries in the form
\be
k'_\mu \,=\, [T_a(k)]_\mu \,=\, T_\mu(a, k), \quad\quad\quad k'_\mu \,=\, [J_\omega(k)]_\mu \,=\,J_\mu(\omega, k),
\ee
where $a$ is a set of four parameters, $\omega$ a set of six parameters, and one has
\be
T_\mu(a, 0) \,=\, a_\mu, \quad\quad\quad J_\mu(\omega, 0) \,=\, 0,
\ee
so that $J_\mu(\omega, k)$ define the six-parameter subgroup of isometries which leave $k=0$ (origin in momentum space) invariant and $T_\mu(a, k)$ are the isometries which do not leave the origin invariant (``translations'').

The proposal to define a RDK is based on the identification of the isometries $k'_\mu = J_\mu(\omega, k)$ with the Lorentz transformation of the total momentum or of one momentum when it does not depend on other momenta (in fact the relativity principle with a deformed composition law of momenta will require the Lorentz transformation of some momenta to depend on other momenta), where $\omega$ are the six parameters of a Lorentz transformation. The dispersion relation will be defined by the points of the hypersurface where a function $C(k)$ takes a constant value (squared mass of the particle). The Lorentz invariance of the dispersion relation reduces to the condition $C(k)=C(k')$ which allows us to determine the function $C$ if one knows the isometries $J_\mu(\omega, k)$ which leave the origin invariant. It is obvious that the dependence on $k$ of $C(k)$ will be the same as that of the distance between the origin and $k$, since this quantity is preserved by isometries that leave the origin invariant.  

Translations $k'_\mu = T_\mu(a, k)$ will be used to define the total momentum $p\oplus q$ of a two-particle system with momenta $p$, $q$ through
\be
(p\oplus q)_\mu \doteq T_\mu(p, q).
\label{DCL-translations}
\ee
It is easy to see that the composition of momenta is then related to the composition of translations, according to
\be
p\oplus q=T_p(q)=T_p(T_q(0))=(T_p \circ T_q)(0).
\label{T-composition}
\ee
Note that the previous equation implies that $T_{(p\oplus q)}$ differs from $(T_p \circ T_q)$ by a transformation that leaves the origin invariant (that is, by a Lorentz transformation).

The two ingredients of a deformed kinematics, a deformed composition law ($\oplus$) and a deformed dispersion relation ($C(k)$), are determined by the isometries $T_a$ and $J_\omega$. Therefore, a derivation of a deformed kinematics from the geometry of momentum space reduces to the identification of the isometries $T_a$, $J_\omega$. The conjecture (proved in Sec.~\ref{sec:diagram}) is that the deformed kinematics obtained in this way, where the deformations of the composition law and dispersion relation are both determined from the isometries of a given metric, is a RDK.

A general study of a RDK from the geometry of momentum space with a given metric $g_{\mu\nu}$ is based on the set of equations
\be
g_{\mu\nu}(T_a(k)) \,=\, \frac{\partial T_\mu(a, k)}{\partial k_\rho} \frac{\partial T_\nu(a, k)}{\partial k_\sigma} g_{\rho\sigma}(k), \quad\quad
g_{\mu\nu}(J_\omega(k)) \,=\, \frac{\partial J_\mu(\omega, k)}{\partial k_\rho} \frac{\partial J_\nu(\omega, k)}{\partial k_\sigma} g_{\rho\sigma}(k),
\label{T,J}
\ee
for the isometries $T_a$, $J_\omega$ which have to be satisfied for any choice of the parameters $a$, $\omega$. 
Taking the limit $k\to 0$ in the set of equations (\ref{T,J}), we get
\be
g_{\mu\nu}(a) \,=\, \left[\lim_{k\to 0} \frac{\partial T_\mu(a, k)}{\partial k_\rho}\right] \, 
\left[\lim_{k\to 0} \frac{\partial T_\nu(a, k)}{\partial k_\sigma}\right] \,\eta_{\rho\sigma}, \quad\quad\quad
\eta_{\mu\nu} \,=\, \left[\lim_{k\to 0} \frac{\partial J_\mu(\omega, k)}{\partial k_\rho}\right] \,  
\left[\lim_{k\to 0} \frac{\partial J_\nu(\omega, k)}{\partial k_\sigma}\right] \,\eta_{\rho\sigma}.
\ee
We can then make the identifications 
\be
\lim_{k\to 0} \frac{\partial T_\mu(a, k)}{\partial k_\rho} \,=\, \delta^\rho_\alpha e_\mu^\alpha(a), \quad\quad\quad
\lim_{k\to 0} \frac{\partial J_\mu(\omega, k)}{\partial k_\rho} \,=\, L_\mu^\rho(\omega),
\label{e,L}
\ee
where $e_\mu^\alpha(k)$ is the (inverse of\footnote{Note that $g_{\mu\nu}$ is the inverse of the metric $g^{\mu\nu}$ in momentum space.} the) tetrad in momentum space, and $L_\mu^\rho(\omega)$ is the standard $(4\times 4)$ matrix representing the Lorentz transformation with parameters $\omega$.
Combining Eq.~\eqref{DCL-translations} with Eq.~\eqref{e,L}, we obtain
\be
\lim_{k\to 0} \frac{\partial(a\oplus k)_\mu}{\partial k_\rho} \,=\, \delta^\rho_\alpha e_\mu^\alpha(a),
\label{magicformula}
\ee
which is a fundamental relationship between (a limit of the derivative of) the composition law and the tetrad in momentum space.

If we consider infinitesimal transformations, we have
\be
T_\mu(\epsilon, k) = k_\mu + \epsilon_\alpha {\cal T}_\mu^\alpha(k), \quad\quad\quad
J_\mu(\epsilon, k) = k_\mu + \epsilon_{\beta\gamma} {\cal J}^{\beta\gamma}_\mu(k),
\ee
and Eq.~(\ref{T,J}) leads to
\be
\frac{\partial g_{\mu\nu}(k)}{\partial k_\rho} {\cal T}^\alpha_\rho(k) \,=\, \frac{\partial{\cal T}^\alpha_\mu(k)}{\partial k_\rho} g_{\rho\nu}(k) +
\frac{\partial{\cal T}^\alpha_\nu(k)}{\partial k_\rho} g_{\mu\rho}(k),
\label{cal(T)}
\ee
\be
\frac{\partial g_{\mu\nu}(k)}{\partial k_\rho} {\cal J}^{\beta\gamma}_\rho(k) \,=\,
\frac{\partial{\cal J}^{\beta\gamma}_\mu(k)}{\partial k_\rho} g_{\rho\nu}(k) +
\frac{\partial{\cal J}^{\beta\gamma}_\nu(k)}{\partial k_\rho} g_{\mu\rho}(k),
\label{cal(J)}
\ee
which is a system of equations for the Killing vectors ${\cal T}^\alpha$, ${\cal J}^{\beta\gamma}$.

Note that if ${\cal T}^\alpha$, ${\cal J}^{\beta\gamma}$ are a solution of the Killing equations (\ref{cal(T)}-\ref{cal(J)}), then ${\cal T}'^\alpha = {\cal T}^\alpha + c^\alpha_{\beta\gamma} {\cal J}^{\beta\gamma}$ is also a solution of Eq.~(\ref{cal(T)}) for any choice of constants $c^\alpha_{\beta\gamma}$, and one has $T'_\mu(\epsilon, 0)=T_\mu(\epsilon, 0)=\epsilon_\mu$. This observation is equivalent to what we commented after Eq.~\eqref{T-composition}. Then, there is an ambiguity in the identification of translations and then on the geometric derivation of a deformed composition law. Writing the generators of isometries as 
\be
T^\alpha \,=\, x^\mu {\cal T}^\alpha_\mu(k), \quad\quad\quad J^{\alpha\beta} \,=\, x^\mu {\cal J}^{\alpha\beta}_\mu(k),
\ee
where $x^\mu$ are the coordinates canonically conjugated to the momenta, $\{p_\mu,x^\nu\}=\delta_\mu^\nu$, we have
\begin{align}
  &\{T^\alpha, T^\beta\} \,=\, x^\rho \left(\frac{\partial{\cal T}^\alpha_\rho(k)}{\partial k_\sigma} {\cal T}^\beta_\sigma(k) - \frac{\partial{\cal T}^\beta_\rho(k)}{\partial k_\sigma} {\cal T}^\alpha_\sigma(k)\right), \\
  &\{T^\alpha, J^{\beta\gamma}\} \,=\, x^\rho \left(\frac{\partial{\cal T}^\alpha_\rho(k)}{\partial k_\sigma} {\cal J}^{\beta\gamma}_\sigma(k) - \frac{\partial{\cal J}^{\beta\gamma}_\rho(k)}{\partial k_\sigma} {\cal T}^\alpha_\sigma(k)\right).
\end{align}
The generators $T^\alpha$, $J^{\beta\gamma}$ will close a Lie algebra since the isometries are a Lie group of transformations. The ambiguity in the identification of translations is just the ambiguity in the choice of basis in the Lie algebra. Each choice of generators of translations  will lead to different deformed composition laws, and then to different relativistic deformed kinematics.

\subsection{Relativistic deformed kinematics}
\label{sec:diagram}

We will now show that the deformed kinematics defined from the isometries of a geometry in momentum space is compatible with the relativity principle (RDK). 
The proof is based on the following diagram:
\begin{center}
\begin{tikzpicture}
\node (v1) at (-2,1) {$q$};
\node (v4) at (2,1) {$\bar q$};
\node (v2) at (-2,-1) {$p \oplus q$};
\node (v3) at (2,-1) {$(p \oplus q)^\prime$};
\draw [->] (v1) edge (v2);
\draw [->] (v4) edge (v3);
\draw [->] (v2) edge (v3);
\node at (-2.6,0) {$T_p$};
\node at (2.7,0) {$T_{p^\prime}$};
\node at (0,-1.4) {$J_\omega$};
\end{tikzpicture}
\end{center}
where the prime denotes the transformation through ${\cal J}_\omega$, and $T_p$, $T_{p'}$ are the translations with parameters $p$ and $p'$. The point $\bar{q}$ is defined as the point whose transformed by the translation $T_{p'}$ is just $(p\oplus q)'$, i.e.,
\be
(p\oplus q)' \,=\, (p' \oplus \bar{q}).
\label{qbar1}
\ee
Note that if one takes $q=0$, then one has $\bar{q}=0$, and in the case $q\neq 0$, the point $\bar{q}$ is obtained from $q$ by a composition of three isometries (the translation $T_p$, a Lorentz transformation $J_\omega$, and the inverse of the translation $T_{p'}$, which will be an isometry since the isometries are a group of transformations). Then there is an isometry which leaves the origin invariant and applied to $q$ gives $\bar{q}$. This means that $q$ and $\bar{q}$ are at the same distance from the origin and then
\be
C(q) \,=\, C(\bar{q}).
\label{qbar2}
\ee
Eqs.~\eqref{qbar1}-\eqref{qbar2} tell us that the deformed kinematics defined by $C$ and $\oplus$ is a relativistic deformed kinematics if we identify the momenta $(p', \bar{q})$ as the Lorentz transformed momenta of $(p, q)$. Eq.~(\ref{qbar1}) implies that the conservation law is Lorentz invariant and Eq.~(\ref{qbar2}), together with $C(p)=C(p')$, that the dispersion relation of the particles is also Lorentz invariant. In the case $p=0$, one has $\bar{q}=q'$, but when $p\neq 0$, $\bar{q}$ will depend on $p$. This means that the Lorentz transformation of two momenta differs from the independent Lorentz transformation of each momentum. The explicit form of this transformation is determined by the composition law and the Lorentz transformation of one momentum. We will show this in an example in the next section.   

\section{Isotropic relativistic deformed kinematics}
\label{sec:examples}

In this section we consider the derivation in detail for two simple cases ($\kappa$-Poincaré and Snyder kinematics) of an isotropic relativistic deformed kinematics and we argue how to proceed when one goes beyond the two simple cases (hybrid models). 

An isotropic relativistic deformed kinematics will be based on an algebra of generators of isometries
\be
\{T^0, T^i\} \,=\, \frac{c_1}{\Lambda} T^i + \frac{c_2}{\Lambda^2} J^{0i}, \quad\quad\quad \{T^i, T^j\} \,=\, \frac{c_2}{\Lambda^2} J^{ij},
\label{isoRDK}
\ee
where we assume that the generators of isometries leaving the origin invariant $J^{\alpha\beta}$ have been chosen to satisfy the standard Lorentz algebra, and the Poisson brackets of $T^\alpha$ and $J^{\beta\gamma}$ are fixed by Jacobi identities.\footnote{The coefficients of $J^{0i}$ and $J^{ij}$ in Eq.~\eqref{isoRDK} are the same also due to Jacobi identities.} For each choice of this algebra (i.e., for each choice of $(c_1/\Lambda)$ and $(c_2/\Lambda^2)$) and for each choice of an isotropic metric one has to find the isometries of the metric whose generators satisfy the chosen algebra. These isometries will define an isotropic relativistic deformed kinematics.

\subsection{$\kappa$-Poincaré relativistic kinematics}

The first simple case one can consider is taking $c_2=0$ in Eq.~\eqref{isoRDK}. The isotropic Lie algebra for the generators of translations will be in this case\footnote{One can reabsorb the coefficient $c_1$ into a redefinition of the scale $\Lambda$.}    
\be
\{T^0, T^i\} \,=\, \pm \frac{1}{\Lambda} T^i.
\label{Talgebra}
\ee
Together with the generators $T^\alpha$ of translations $k \to T_a(k) = (a\oplus k)$, one can consider the generators $\tilde{T}^\alpha$ of transformations $k \to (k\oplus a)$, which can be identified from (\ref{e,L}),
\be
\tilde{T}^\alpha \,=\, x^\mu e^\alpha_\mu(k).
\label{Ttilde}
\ee
A result of differential geometry~\cite{Chern:1999jn} is that, when the generators of translations $T^\alpha$ satisfy the Lie algebra (\ref{Talgebra}), then $\tilde{T}^\alpha$ are also the generators of a Lie algebra, 
\be
 \{\tilde{T}^0, \tilde{T}^i\} \,=\, \mp \frac{1}{\Lambda} \tilde{T}^i ,
\label{Ttildealgebra}
\ee   
which is the Lie algebra of the generators of translations up to a sign.\footnote{The fact that the directional tetrad derivatives ($\tilde{T}^\alpha$) close the same Lie algebra of the translation generators ($T^\alpha$), but with different sign, is also understood from the identification of $\tilde{T}^\alpha$ and $T^\alpha$ as right- and left-translation generators.} Both algebras \eqref{Talgebra}-\eqref{Ttildealgebra} are just the algebra for the coordinates of $\kappa$-Minkowski spacetime.  The problem to determine a tetrad $e^\alpha_\mu(k)$ compatible with the algebra (\ref{Ttildealgebra}) reduces then to the problem of finding a representation of $\kappa$-Minkowski spacetime coordinates in terms of canonical phase space coordinates. One can easily check that a possible choice for $\kappa$-Minkowski spacetime coordinates, and then for the tetrad, is
\be
e^0_0(k) \,=\, 1, \quad\quad\quad e^0_i(k) \,=\, e^i_0(k) \,=\, 0, \quad\quad\quad e^i_j (k) \,=\, \delta^i_j e^{\mp k_0/\Lambda}.
\label{bicross-tetrad}
\ee

We now have to define the finite translations $T_\mu(a,k)$ whose generators satisfy Eq.~\eqref{Talgebra}. Since this is a Lie algebra, the translation transformations must form a group, that is, the translations will be a subgroup of the isometries of the metric. Inspired by the first equation in Eq.~\eqref{e,L}, we try to define the finite translation as a transformation which leaves the tetrad invariant:
\be
e_\mu^\alpha(T(a, k)) \,=\, \frac{\partial T_\mu(a, k)}{\partial k_\nu} \,e_\nu^\alpha(k).
\label{T(a,k)}
\ee
It is evident that if $T_\mu(a,k)$ is a solution to this equation, which implies that a translation leaves the tetrad, and therefore, the metric, invariant (so that it is indeed an isometry), then translations form a group, since the composition of two transformations that leave the tetrad invariant has also the same property. Indeed, Eq. \eqref{T(a,k)} can be solved, and this allows to get the finite translations.
The solution of (\ref{T(a,k)}) for $T_\mu(a, k)$ is
\be
T_0(a, k) \,=\, a_0 + k_0, \quad\quad\quad T_i(a, k) \,=\, a_i + k_i e^{\mp a_0/\Lambda},
\ee
and the corresponding composition law of momenta 
\be
(p\oplus q)_0 \,=\, T_0(p, q) \,=\, p_0 + q_0, \quad\quad\quad
(p\oplus q)_i \,=\, T_i(p, q) \,=\, p_i + q_i e^{\mp p_0/\Lambda},
\label{kappa-dcl}
\ee
is just the composition law of momenta of $\kappa$-Poincaré kinematics in the bicrossproduct basis, or a related composition law obtained by replacing $\Lambda$ by $-\Lambda$.

The function $C(k)$ invariant under isometries which leave the origin invariant will satisfy the equation
\be
\frac{\partial C(k)}{\partial k_\mu} \,{\cal J}^{\alpha\beta}_\mu(k) \,=\, 0 ,
\ee
where ${\cal J}^{\alpha\beta}$ are the Killing vectors satisfying Eq.~\eqref{cal(J)} with the metric $g_{\mu\nu}(k)=e^\alpha_\mu(k)\eta_{\alpha\beta}e^\beta_\nu(k)$ defined by the tetrad~\eqref{bicross-tetrad}:
\be
0 \,=\, \frac{{\cal J}^{\alpha\beta}_0(k)}{\partial k_0}, \quad\quad
0 \,=\, - \frac{{\cal J}^{\alpha\beta}_0(k)}{\partial k_i} e^{\mp 2k_0/\Lambda} + \frac{{\cal J}^{\alpha\beta}_i(k)}{\partial k_0}, \quad\quad
\pm \frac{2}{\Lambda} {\cal J}^{\alpha\beta}_0(k) \delta_{ij} \,=\, - \frac{\partial{\cal J}^{\alpha\beta}_i(k)}{\partial k_j} - \frac{\partial{\cal J}^{\alpha\beta}_j(k)}{\partial k_i} .
\ee
The solution for ${\cal J}^{\alpha\beta}_\mu(k)$ is
 \be
{\cal J}^{0i}_0(k) \,=\, -k_i, \quad \quad \quad {\cal J}^{0i}_j(k)\,=\, \pm \delta^i_j \,\frac{\Lambda}{2} \left[e^{\mp 2 k_0/\Lambda} - 1 - \frac{\vec{k}^2}{\Lambda^2}\right] \pm \,\frac{k_i k_j}{\Lambda},
\ee 
and then 
\be
C(k) \,=\, \Lambda^2 \left(e^{k_0/\Lambda} + e^{-k_0/\Lambda} - 2\right) - e^{\pm k_0/\Lambda} \vec{k}^2  \,,
\ee
which is the function of the momentum defining the dispersion relation of $\kappa$-Poincaré kinematics in the bicrossproduct basis (or a related dispersion relation obtained by replacing $\Lambda$ by $-\Lambda$).

Once the composition law of momenta (\ref{kappa-dcl}) and the (infinitesimal) Lorentz transformation of momenta
\be
k'_\mu = k_\mu + \epsilon_{\alpha\beta} {\cal J}_\mu^{\alpha\beta}(k)
\ee
have been determined, we can go back to the diagram in Sec.~\ref{sec:diagram} and determine the explicit form of $\bar{q}$ for an infinitesimal Lorentz transformation from the relation
\be
(p\oplus q)' \,=\, p'\oplus \bar{q},
\ee
by equating the terms linear in $\epsilon_{\alpha\beta}$ in the difference from $p\oplus q$ on both sides of this equation. Then one finds
\be
\epsilon_{\alpha\beta} {\cal J}^{\alpha\beta}_\mu(p\oplus q) \,=\, \epsilon_{\alpha\beta} \frac{\partial(p\oplus q)_\mu}{\partial p_\nu} {\cal J}^{\alpha\beta}_\nu(p) + \frac{\partial(p\oplus q)_\mu}{\partial q_\nu} (\bar{q}_\nu - q_\nu).
\ee
From the composition law (\ref{kappa-dcl}) with the minus sign in the exponent, we have
\begin{align}
& \frac{\partial(p\oplus q)_0}{\partial p_0} \,=\, 1, \quad\quad
\frac{\partial(p\oplus q)_0}{\partial p_i} \,=\, 0, \quad\quad
\frac{\partial(p\oplus q)_i}{\partial p_0} \,=\, - \frac{q_i}{\Lambda} e^{-p_0/\Lambda}, \quad\quad
\frac{\partial(p\oplus q)_i}{\partial p_j} \,=\, \delta_i^j, \\
& \frac{\partial(p\oplus q)_0}{\partial q_0} \,=\, 1, \quad\quad  \frac{\partial(p\oplus q)_0}{\partial q_i} \,=\, 0, \quad\quad
\frac{\partial(p\oplus q)_i}{\partial q_0} \,=\, 0, \quad\quad  \frac{\partial(p\oplus q)_i}{\partial q_j} \,=\, \delta_i^j e^{-p_0/\Lambda}.
\end{align}
Then we find
\be
\bar{q}_0 \,=\, q_0 + \epsilon_{\alpha\beta} \left[{\cal J}^{\alpha\beta}_0(p\oplus q) - {\cal J}^{\alpha\beta}_0(p)\right], \quad\quad\quad
\bar{q}_i \,=\, q_i + \epsilon_{\alpha\beta} \, e^{p_0/\Lambda} \, \left[{\cal J}^{\alpha\beta}_i(p\oplus q) - {\cal J}^{\alpha\beta}_i(p) + \frac{q_i}{\Lambda} e^{-p_0/\Lambda} {\cal J}^{\alpha\beta}_0(p)\right],
\ee
and one can check that the Lorentz transformation of two momenta $(p, q) \to (p', \bar{q})$ reproduces the coproduct of Lorentz generators of $\kappa$-Poincaré in the bicrossproduct basis. 

We have seen then that the $\kappa$-Poincaré kinematics in the bicrossproduct basis is just the deformed kinematics obtained from a de Sitter momentum space when one uses coordinates such that the tetrad takes the form in (\ref{bicross-tetrad}) and one defines the translations as the isometries which leave the tetrad invariant. Different choices of tetrad, corresponding to different representations of $\kappa$-Minkowski spacetime coordinates in terms of canonical phase space variables, will lead to $\kappa$-Poincaré kinematics in different basis, that is, they will correspond to different choices of coordinates in momentum space of the deformed kinematics defined by the algebra~\eqref{Talgebra}.

The momentum space metric for the tetrad in (\ref{bicross-tetrad}) is~\footnote{This is the metric in the comoving coordinate system of de Sitter space~\cite{Gubitosi:2013rna}.}
\be
g_{00}(k) \,=\, 1, \quad\quad\quad g_{0i}(k) \,=\, g_{i0}(k) \,=\, 0, \quad\quad\quad g_{ij}(k) \,=\, - \delta_{ij} e^{\mp 2k_0/\Lambda}.
\label{bicross-metric}
\ee
One can check from the calculation of the Riemann-Christoffel tensor that this is the metric of a de Sitter momentum space with curvature $(12/\Lambda^2)$.\footnote{In Appendix~\ref{apendice} we show why a deformed kinematics based on the invariance of the tetrad can not be found in the case of anti-de Sitter momentum space.} 

In summary, the geometric interpretation of the deformed kinematics allowed us to produce the results of the $\kappa$-Poincaré Hopf algebra
in the bicrossproduct basis~\cite{KowalskiGlikman:2002we}. In the algebraic approach, one imposes that the spacetime coordinates form a closed subalgebra, and a closed algebra with the generators $J^{\alpha\beta}$. These requirements are automatically satisfied in the geometric approach, where the spacetime coordinates are just the generators of a subgroup of isometries (translations). This also implies that the composition law is associative. Note also that, if translations are a subgroup of isometries, the composition law is associative.

\subsection{Beyond $\kappa$-Poincaré relativistic kinematics}

Another simple choice for the constants in the algebra of the generators of translations is $c_1=0$, which gives a covariant algebra, the so-called Snyder algebra\cite{Snyder:1946qz}. If one goes back to the simplest covariant form for the metric of de Sitter momentum space, $g_{\mu\nu}(k) =\eta_{\mu\nu} + k_\mu k_\nu/\Lambda^2$, the invariance of the metric under translations, together with the definition of the composition law of momenta from translations, leads to the system of equations for the composition of momenta
\be
\eta_{\mu\nu} + \frac{(p\oplus q)_\mu (p\oplus q)_\nu}{\Lambda^2} \,=\, \frac{\partial(p\oplus q)_\mu}{\partial q_\rho} \frac{\partial(p\oplus q)_\nu}{\partial q_\sigma} \left(\eta_{\rho\sigma} + \frac{q_\rho q_\sigma}{\Lambda^2}\right).
\ee
Using the general form of a covariant composition of momenta
\be
(p\oplus q)_\mu \,= p_\mu f_L\left(p^2/\Lambda^2, p.q/\Lambda^2, q^2/\Lambda^2\right) + q_\mu f_R\left(p^2/\Lambda^2, p.q/\Lambda^2, q^2/\Lambda^2\right),
\label{DCLSnyder-1}
\ee
we have a system of equations for the two functions $f_L$, $f_R$ of three variables. The solution of the system of equations is
\be
f_L\left(p^2/\Lambda^2, p.q/\Lambda^2, q^2/\Lambda^2\right) \,=\, \sqrt{1+\frac{q^2}{\Lambda^2}}+\frac{p\cdot q}{\Lambda^2\left(1+\sqrt{1+p^2/\Lambda^2}\right)}   ,\quad\quad\quad
f_R\left(p^2/\Lambda^2, p.q/\Lambda^2, q^2/\Lambda^2\right) \,=\, 1,
\label{DCLSnyder-2}
\ee
and one reproduces the composition of momenta of Snyder kinematics in the Maggiore basis (it is the complete version of the first orders obtained in Ref.~\cite{Banburski:2013jfa}).

If we obtain the infinitesimal generators
\be
{\cal T}^\mu_\nu(p)=\left.\frac{\partial \left(k\oplus p\right)_\nu}{\partial k_\mu} \right\rvert_{k \rightarrow 0}\,=\,\delta^\mu_\nu \sqrt{1+\frac{p^2}{\Lambda^2}} \,,
\ee  
one can check that $T^\alpha=x^\nu{\cal T}^\alpha_\nu$ satisfies
\be
\{T^\alpha, T^\beta\} \,=\, \frac{1}{\Lambda^2} J^{\alpha\beta},
\ee
which is indeed the Snyder algebra. Covariance of the algebra also tells us that $C(p)$ is a function of $p^2$, and that the Lorentz transformation in the two-particle system is linear, that is, it is the same than that in the one-particle system.

For other choices of coordinates with a covariant metric one would find the Snyder kinematics in different basis. All the discussion of a deformed covariant metric can be repeated in the case of anti-de Sitter momentum space and one will obtain the same results replacing $(1/\Lambda^2)$ by $-(1/\Lambda^2)$.   

In the case where both coefficients $c_1$, $c_2$ are non-zero, one has a relativistic deformed kinematics depending on one additional dimensionless parameter which interpolates between Snyder and $\kappa$-Poincaré kinematics. One has a system of equations for the translation $T_a(p)$ (and then for the deformed composition law of momenta)  when one considers a given algebra of isometries and the translation invariance of a choice of de Sitter or anti-de Sitter metric. When one considers the general expression for an isotropic composition of momenta as an expansion in powers of $(1/\Lambda)$, one finds at each order in this expansion a set of equations which determine the dimensionless coefficients which define the power expansion of the deformed composition of momenta. This will reproduce the results of the kinematics of what is known as hybrid models~\cite{Meljanac:2009ej}.

We note that while the composition law of $\kappa$-Poincar\'e kinematics is associative, this is no longer the case for Snyder and hybrid kinematics.\footnote{One can check explicitly, using Eqs.~\eqref{DCLSnyder-1}-\eqref{DCLSnyder-2}, that $k\oplus(p\oplus q) \neq (k\oplus p)\oplus q$.} In fact, a generic composition law of momenta (defined as a map from a pair of momentum variables into a momentum variable such that when one of the momentum variables in the pair is zero reduces no the non-zero momentum variable) will not be associative. The associativity of the composition law of $\kappa$-Poincar\'e kinematics is a direct consequence of the property that the generators $T^\alpha$ of the isometries which define in the geometric approach the composition law, define a subalgebra of the Lie algebra of isometries. In the case of Snyder or hybrid kinematics, the generators $T^\alpha$ do not define a subalgebra ($c_2\neq 0$ in Eq.~\eqref{isoRDK}) and the corresponding composition law is not associative.

The composition law of $\kappa$-Poincar\'e kinematics obtained in the geometric approach can also be derived from the coproduct of the $\kappa$-Poincar\'e Hopf algebra, and the associativity of the composition law is just a consequence of the coassociativity of the coproduct of the Hopf algebra. If one generalizes the correspondence between the composition law and the coproduct of $\kappa$-Poincar\'e to the composition law of Snyder kinematics, then one obtains a coproduct which is not coassociative as a consequence of the non-associativity of the composition law. This is the reason why, if one wants to obtain Snyder kinematics in the algebraic approach, one has to go beyond the framework of Hopf algebras~\cite{Battisti:2010sr}.

The restriction on the kinematics derived from a Hopf algebra (associativity of the composition law due to the coassociativity of the coproduct) can be also obtained in the geometric approach if one imposes the condition that the composition law should define a group of transformations in momentum space. Since associativity is a crucial property to simplify the analysis of processes in the deformed kinematics, we will stick to this very important particular case, in which it is also possible to make a comparison between our results and previous studies of the geometric approach. This will be the subject of the next Section.

\section{Comparison with previous attempts to relate a deformation of the kinematics with the geometry in momentum space}
\label{relative_locality}

In previous attempts to establish a relation between a deformation of the kinematics and a geometry of momentum space~\cite{AmelinoCamelia:2011bm}, the basic ingredients have been the identification of the function of momenta which defines the dispersion relation with the square of the distance from the origin to a point in momentum space in terms of its coordinates, and a relation between the deformed composition law and a (non-metrical) connection in momentum space. 
Specifically, Ref.~\cite{AmelinoCamelia:2011bm} defines the value of the connection at the origin as
\be
\Gamma^{\tau \lambda}_\nu (0)\,=\,-\left.\frac{\partial^2  (p\oplus q)_\nu}{\partial p_\tau \partial q_\lambda}\right\rvert_{p,q \rightarrow 0}\,.
\ee
In order to determine the connection away from the origin, the authors use a prescription based on a $k$-dependent momentum composition
\be
(p\oplus_k q) \,\doteq\, k\oplus\left((\hat{k}\oplus p)\oplus(\hat{k}\oplus q)\right),
\label{k-DCL}
\ee
where $\hat{k}$ is called the \emph{antipode} of $k$, defined as
\be
k\oplus \hat{k}=\hat{k}\oplus k =0.
\ee
This allows them to define
\be
\Gamma^{\tau \lambda}_\nu (k)\,=\,-\left.\frac{\partial^2  (p\oplus_{k}q)_\nu}{\partial p_\tau \partial q_\lambda}\right\rvert_{p,q \rightarrow k}\,.
\label{k-connection}
\ee

In this work, however, we have completed the proposal of Ref.~\cite{Lobo:2016blj}, defining the deformed composition law through some isometries (translations) of the metric, without any reference to a connection. Which is the relationship between our approach and the geometric approach of Ref.~\cite{AmelinoCamelia:2011bm}? We will now show how the definition of the DCL from translations which leave the tetrad (and then the metric) invariant leads to identify a combination of the tetrad and its derivatives which transforms as a connection (but in general differs from the metric connection). This connection will allow us to compare the direct simple derivation of a RDK from the geometry presented in this work with previous approaches based on a relation between a connection and a deformed composition law. 

We start by writing the equation of invariance of the tetrad under translations (Eq.~\eqref{T(a,k)}) in terms of the composition law, and taking the derivative on both sides with respect to $p_\tau$
\be
\frac{\partial e^\alpha_\nu(p\oplus q)}{\partial p_\tau} \,=\, \frac{\partial  e^\alpha_\nu(p\oplus q)}{\partial (p\oplus q)_\sigma} \frac{\partial(p\oplus q)_\sigma}{\partial p_\tau} \,=\, \frac{\partial^2(p\oplus q)_\nu}{\partial p_\tau \partial q_\rho} e^\alpha_\rho(q).
\ee 
From this we can get the second derivative of the composition law
\be
\frac{\partial^2(p\oplus q)_\nu}{\partial p_\tau \partial q_\rho} \,=\, e^\rho_\alpha(q) \frac{\partial  e^\alpha_\nu(p\oplus q)}{\partial (p\oplus q)_\sigma} \frac{\partial(p\oplus q)_\sigma}{\partial p_\tau},
\ee
where $e^\rho_\alpha$ is the inverse of $e^\alpha_\rho$, $e^\alpha_\nu e^\mu_\alpha=\delta^\mu_\nu$. 
But also using Eq.~\eqref{T(a,k)}, one has
\be
e^\rho_\alpha(q) \,=\, \frac{\partial(p\oplus q)_\mu}{\partial q_\rho} e^\mu_\alpha(p\oplus q)
\label{magicformula2}
\ee
and then 
\be
\frac{\partial^2(p\oplus q)_\nu}{\partial p_\tau \partial q_\rho} + \Gamma^{\sigma\mu}_\nu(p\oplus q) \,\frac{\partial(p\oplus q)_\sigma}{\partial p_\tau} \,\frac{\partial(p\oplus q)_\mu}{\partial q_\rho} \,=\, 0,
\label{geodesic_tetrad}
\ee
where 
\be
\Gamma^{\sigma\mu}_\nu(k) \,\doteq\, - e^\mu_\alpha(k) \, \frac{\partial  e^\alpha_\nu(k)}{\partial k_\sigma}.
\label{e-connection}
\ee
One can easily check that the combination of the tetrad and derivatives in Eq.~\eqref{e-connection} transforms like a connection does under a change of coordinates in momentum space.

In Ref.~\cite{Amelino-Camelia:2013sba}, a way to introduce a connection in momentum space and define the deformed composition through the parallel transport defined by the connection was proposed as the link between the geometry of momentum space and a DCL. It turns out that the composition law defined in this way satisfies Eq.~\eqref{geodesic_tetrad}. This equation does not allow to determine the composition law for a given connection unless we add the condition of associativity for the composition law. Then we have a proof that the composition law defined by a translation which leaves a tetrad invariant is the associative composition law derived by parallel transport, with the connection (\ref{e-connection}) associated to the tetrad.

Finally, if the composition law is associative, then Eq.~\eqref{k-DCL} reduces to
\be
(p\oplus_k q) \,=\, p\oplus\hat{k}\oplus q.
\ee
We can then get a differential equation for this $k$-dependent composition law if we replace $q$ by $(\hat{k}\oplus q)$ in Eq.~\eqref{geodesic_tetrad}, which is valid for any momenta ($p, q$). We have
\be
\frac{\partial^2  (p \oplus \hat{k} \oplus q)_\nu}{\partial p_\tau \partial(\hat{k} \oplus q)_\rho}+\Gamma^{\sigma \mu}_\nu (p \oplus \hat{k} \oplus q) \frac{\partial (p \oplus \hat{k} \oplus q)_\sigma}{\partial p_\tau}\frac{\partial (p \oplus \hat{k} \oplus q)_\mu}{\partial(\hat{k} \oplus q)_\rho}\,=\,0\,.
\ee
Now, multiplying by $\partial(\hat{k} \oplus q)_\rho/\partial q_\lambda$, one obtains
\be
\frac{\partial^2  (p \oplus \hat{k} \oplus q)_\nu}{\partial p_\tau \partial q_\lambda}+\Gamma^{\sigma \mu}_\nu (p \oplus \hat{k} \oplus q) \frac{\partial (p \oplus \hat{k} \oplus q)_\sigma}{\partial p_\tau}\frac{\partial (p \oplus \hat{k} \oplus q)_\mu}{\partial q_\lambda}\,=\,0\,.
\label{connection_1}
\ee
Taking $p=q=k$ in Eq.~\eqref{connection_1}, we get
\be
\Gamma^{\tau \lambda}_\nu (k)\,=\,-\left.\frac{\partial^2  (p\oplus_{k}q)_\nu}{\partial p_\tau \partial q_\lambda}\right\rvert_{p,q \rightarrow k}\,,
\ee
which is the expression used in Ref.~\cite{AmelinoCamelia:2011bm}, Eq.~\eqref{k-connection}.
This proofs that the connection associated to the tetrad (\ref{e-connection}) in our approach is related to the composition law of momenta defined by the invariance under translations of the tetrad by the prescription proposed in Ref.~\cite{AmelinoCamelia:2011bm} to define a non-metric connection from a composition law of momenta.   

\section{Summary and outlook}
\label{conclusions}

We have presented in this work a proposal to derive a (relativistic) deformed kinematics from a geometry in momentum space defining a deformed composition law from the translations in momentum space and the deformed dispersion relation from the (square of the) distance from the origin to a point in momentum space. In the case of a maximally symmetric space one gets a relativistic deformed kinematics.

A comparison with other attempts to relate a geometry with a deformed kinematics based on a relation between a (non-metric) connection in momentum space and a deformed composition of momenta has been discussed. In fact we have identified a connection in our proposal that, in the case of an associative composition of momenta, satisfies the relations with the deformed composition law proposed in those attempts. This connection, however, is derived from the tetrad, which in our approach is a more fundamental object that relates the geometry and the deformed composition law through Eq.~\eqref{magicformula}.

We have discussed the systematic method to determine a relativistic isotropic deformed kinematics for a given Lie algebra of generators of isometries, and
compared this geometric approach with the standard algebraic approach where a curved momentum space is deduced from a quantum-deformed Hopf algebra. The algebraic approach is reproduced when the generators of translations close an algebra, Eq.~\eqref{Talgebra}. This condition allows one to completely determine the composition law, since Eq.~\eqref{magicformula} is converted into Eq.~\eqref{magicformula2}. We have seen that in this case, a particular choice of 
de Sitter momentum space coordinates reproduces the Hopf algebra  structure of coproduct of momenta and boosts, as well as the Casimir, of $\kappa$-Poincaré in the bicrossproduct basis, which are then obtained in a simple and intuitive way.

The geometric approach also includes other cases. For a linear Lorentz covariant metric, the composition law of momenta is obtained for the Snyder noncommutativity in the Maggiore representation. The details of the derivation of the kinematics of hybrid models, as well as the possibility to consider anisotropic relativistic deformed kinematics, will be the subject of a future work.

Finally, let us mention that the consistency of the independence of the geometry on the choice of coordinates (momentum variables) with the identification of physical momentum variables (which is implicit in any study of the phenomenological consequences of a deformation of the kinematics) is an open problem to be studied, as well as the possibility to extend the derivation of a (not relativistic) deformed kinematics from a geometry in momentum space beyond the maximally symmetric cases.

\appendix
\section{Algebra of isometry generators in de Sitter and anti-de Sitter spaces}
\label{apendice}

We start from a choice of generators of isometries in de Sitter space $(T^\alpha_S, J^{\beta\gamma})$ such that the Lie algebra of isometries is Lorentz covariant
\be
\{T^\alpha_S, T^\beta_S\} \,=\, \frac{J^{\alpha\beta}}{\Lambda^2}, \quad\quad \{T^\alpha_S, J^{\beta\gamma}\} \,=\, \eta^{\alpha\beta} T^\gamma_S - \eta^{\alpha\gamma} T^\beta_S, \quad\quad \lbrace J^{\alpha\beta},J^{\gamma\delta}\rbrace\,=\, \eta^{\beta\gamma}J^{\alpha\delta} - \eta^{\alpha\gamma}J^{\beta\delta} - \eta^{\beta\delta}J^{\alpha\gamma} + \eta^{\alpha\delta}J^{\beta\gamma}.
\label{cov_generators}
\ee
If we want to have a four dimensional subalgebra (maintaining isotropy) we have to change the basis of three translation generators
\be
T^0_\kappa \,=\, T^0_S, \quad\quad\quad  T^i_\kappa \,=\, T^i_S \pm \frac{J^{0i}}{\Lambda},
\label{eq:change_generators}
\ee
so that the new translation generators satisfy the closed algebra
\be
\lbrace T_\kappa^0, T_\kappa^i\rbrace\,=\, \mp \frac{T_\kappa^i}{\Lambda}\,.
\ee
This is the algebra of isometries that leads to $\kappa$-Poincar\'e kinematics, as we show in Sec.~\ref{sec:examples}. 

If one tries to make a change of variables similar to Eq.~\eqref{eq:change_generators} starting from the Lorentz covariant Lie algebra of isometries in anti-de Sitter space (replacing $\Lambda^2$ by $-\Lambda^2$ in (\ref{cov_generators})), one finds that it is not possible to find a change of basis for translation generators so that they close a four dimensional algebra. Then there is no isotropic deformed relativistic deformed kinematics with an associative deformed composition law in the case of anti-de Sitter momentum space.

The most general choice of translation generators (maintaining isotropy) is
\be
T^0_H \,=\, T^0_S, \quad\quad\quad T^i_H \,=\, T^i_S + \alpha \frac{J^{0i}}{\Lambda},
\ee
with a new arbitrary parameter $\alpha$. The corresponding deformed composition law will reproduce the kinematics of hybrids models which interpolate between the covariant Snyder kinematics and $\kappa$-Poincar\'e kinematics.

\section*{Acknowledgments}
This work is supported by the Spanish grants FPA2015-65745-P (MINECO/FEDER), PGC2018-095328-B-I00 (FEDER/Agencia estatal de investigación), and by the Spanish DGIID-DGA Grant No. 2015-E24/2.
The authors would also like to thank support from the COST Action CA18108, and acknowledge César Asensio, Jesús Clemente, Ángel Ballesteros and Flavio Mercati for useful conversations.


\begin{thebibliography}{29}%
\makeatletter
\providecommand \@ifxundefined [1]{%
 \@ifx{#1\undefined}
}%
\providecommand \@ifnum [1]{%
 \ifnum #1\expandafter \@firstoftwo
 \else \expandafter \@secondoftwo
 \fi
}%
\providecommand \@ifx [1]{%
 \ifx #1\expandafter \@firstoftwo
 \else \expandafter \@secondoftwo
 \fi
}%
\providecommand \natexlab [1]{#1}%
\providecommand \enquote  [1]{``#1''}%
\providecommand \bibnamefont  [1]{#1}%
\providecommand \bibfnamefont [1]{#1}%
\providecommand \citenamefont [1]{#1}%
\providecommand \href@noop [0]{\@secondoftwo}%
\providecommand \href [0]{\begingroup \@sanitize@url \@href}%
\providecommand \@href[1]{\@@startlink{#1}\@@href}%
\providecommand \@@href[1]{\endgroup#1\@@endlink}%
\providecommand \@sanitize@url [0]{\catcode `\\12\catcode `\$12\catcode
  `\&12\catcode `\#12\catcode `\^12\catcode `\_12\catcode `\%12\relax}%
\providecommand \@@startlink[1]{}%
\providecommand \@@endlink[0]{}%
\providecommand \url  [0]{\begingroup\@sanitize@url \@url }%
\providecommand \@url [1]{\endgroup\@href {#1}{\urlprefix }}%
\providecommand \urlprefix  [0]{URL }%
\providecommand \Eprint [0]{\href }%
\providecommand \doibase [0]{http://dx.doi.org/}%
\providecommand \selectlanguage [0]{\@gobble}%
\providecommand \bibinfo  [0]{\@secondoftwo}%
\providecommand \bibfield  [0]{\@secondoftwo}%
\providecommand \translation [1]{[#1]}%
\providecommand \BibitemOpen [0]{}%
\providecommand \bibitemStop [0]{}%
\providecommand \bibitemNoStop [0]{.\EOS\space}%
\providecommand \EOS [0]{\spacefactor3000\relax}%
\providecommand \BibitemShut  [1]{\csname bibitem#1\endcsname}%
\let\auto@bib@innerbib\@empty
\bibitem [{\citenamefont {Mukhi}(2011)}]{Mukhi:2011zz}%
  \BibitemOpen
  \bibfield  {author} {\bibinfo {author} {\bibfnamefont {S.}~\bibnamefont
  {Mukhi}},\ }\href {\doibase 10.1088/0264-9381/28/15/153001} {\bibfield
  {journal} {\bibinfo  {journal} {Class. Quant. Grav.}\ }\textbf {\bibinfo
  {volume} {28}},\ \bibinfo {pages} {153001} (\bibinfo {year} {2011})},\
  \Eprint {http://arxiv.org/abs/1110.2569} {arXiv:1110.2569 [physics.pop-ph]}
  \BibitemShut {NoStop}%
\bibitem [{\citenamefont {Aharony}(2000)}]{Aharony:1999ks}%
  \BibitemOpen
  \bibfield  {author} {\bibinfo {author} {\bibfnamefont {O.}~\bibnamefont
  {Aharony}},\ }\bibfield  {booktitle} {\emph {\bibinfo {booktitle} {{Strings
  '99. Proceedings, Conference, Potsdam, Germany, July 19-24, 1999}}},\ }\href
  {\doibase 10.1088/0264-9381/17/5/302} {\bibfield  {journal} {\bibinfo
  {journal} {Class. Quant. Grav.}\ }\textbf {\bibinfo {volume} {17}},\ \bibinfo
  {pages} {929} (\bibinfo {year} {2000})},\ \Eprint
  {http://arxiv.org/abs/hep-th/9911147} {arXiv:hep-th/9911147 [hep-th]}
  \BibitemShut {NoStop}%
\bibitem [{\citenamefont {Dienes}(1997)}]{Dienes:1996du}%
  \BibitemOpen
  \bibfield  {author} {\bibinfo {author} {\bibfnamefont {K.~R.}\ \bibnamefont
  {Dienes}},\ }\bibfield  {booktitle} {\emph {\bibinfo {booktitle} {{Institute
  for Theoretical Physics Conference on Unification: From the Weak Scale to the
  Planck Scale Santa Barbara, California, October 23-27, 1995}}},\ }\href
  {\doibase 10.1016/S0370-1573(97)00009-4} {\bibfield  {journal} {\bibinfo
  {journal} {Phys. Rept.}\ }\textbf {\bibinfo {volume} {287}},\ \bibinfo
  {pages} {447} (\bibinfo {year} {1997})},\ \Eprint
  {http://arxiv.org/abs/hep-th/9602045} {arXiv:hep-th/9602045 [hep-th]}
  \BibitemShut {NoStop}%
\bibitem [{\citenamefont {Sahlmann}(2010)}]{Sahlmann:2010zf}%
  \BibitemOpen
  \bibfield  {author} {\bibinfo {author} {\bibfnamefont {H.}~\bibnamefont
  {Sahlmann}},\ }in\ \href
  {https://inspirehep.net/record/843661/files/arXiv:1001.4188.pdf} {\emph
  {\bibinfo {booktitle} {{Proceedings, Foundations of Space and Time:
  Reflections on Quantum Gravity: Cape Town, South Africa}}}}\ (\bibinfo {year}
  {2010})\ pp.\ \bibinfo {pages} {185--210},\ \Eprint
  {http://arxiv.org/abs/1001.4188} {arXiv:1001.4188 [gr-qc]} \BibitemShut
  {NoStop}%
\bibitem [{\citenamefont {Dupuis}\ \emph {et~al.}(2012)\citenamefont {Dupuis},
  \citenamefont {Ryan},\ and\ \citenamefont {Speziale}}]{Dupuis:2012yw}%
  \BibitemOpen
  \bibfield  {author} {\bibinfo {author} {\bibfnamefont {M.}~\bibnamefont
  {Dupuis}}, \bibinfo {author} {\bibfnamefont {J.~P.}\ \bibnamefont {Ryan}}, \
  and\ \bibinfo {author} {\bibfnamefont {S.}~\bibnamefont {Speziale}},\ }\href
  {\doibase 10.3842/SIGMA.2012.052} {\bibfield  {journal} {\bibinfo  {journal}
  {SIGMA}\ }\textbf {\bibinfo {volume} {8}},\ \bibinfo {pages} {052} (\bibinfo
  {year} {2012})},\ \Eprint {http://arxiv.org/abs/1204.5394} {arXiv:1204.5394
  [gr-qc]} \BibitemShut {NoStop}%
\bibitem [{\citenamefont {Van~Nieuwenhuizen}(1981)}]{VanNieuwenhuizen:1981ae}%
  \BibitemOpen
  \bibfield  {author} {\bibinfo {author} {\bibfnamefont {P.}~\bibnamefont
  {Van~Nieuwenhuizen}},\ }\href {\doibase 10.1016/0370-1573(81)90157-5}
  {\bibfield  {journal} {\bibinfo  {journal} {Phys. Rept.}\ }\textbf {\bibinfo
  {volume} {68}},\ \bibinfo {pages} {189} (\bibinfo {year} {1981})}\BibitemShut
  {NoStop}%
\bibitem [{\citenamefont {Taylor}(1984)}]{Taylor:1983su}%
  \BibitemOpen
  \bibfield  {author} {\bibinfo {author} {\bibfnamefont {J.~G.}\ \bibnamefont
  {Taylor}},\ }\href {\doibase 10.1016/0146-6410(84)90002-4} {\bibfield
  {journal} {\bibinfo  {journal} {Prog. Part. Nucl. Phys.}\ }\textbf {\bibinfo
  {volume} {12}},\ \bibinfo {pages} {1} (\bibinfo {year} {1984})}\BibitemShut
  {NoStop}%
\bibitem [{\citenamefont {Wallden}(2010)}]{Wallden:2010sh}%
  \BibitemOpen
  \bibfield  {author} {\bibinfo {author} {\bibfnamefont {P.}~\bibnamefont
  {Wallden}},\ }\bibfield  {booktitle} {\emph {\bibinfo {booktitle} {{Classical
  and quantum gravity. Proceedings, 1st Mediterranean Conference, MCCQG 2009,
  Kolymbari, Crete, Greece, September 14-18, 2009}}},\ }\href {\doibase
  10.1088/1742-6596/222/1/012053} {\bibfield  {journal} {\bibinfo  {journal}
  {J. Phys. Conf. Ser.}\ }\textbf {\bibinfo {volume} {222}},\ \bibinfo {pages}
  {012053} (\bibinfo {year} {2010})},\ \Eprint {http://arxiv.org/abs/1001.4041}
  {arXiv:1001.4041 [gr-qc]} \BibitemShut {NoStop}%
\bibitem [{\citenamefont {Wallden}(2013)}]{Wallden:2013kka}%
  \BibitemOpen
  \bibfield  {author} {\bibinfo {author} {\bibfnamefont {P.}~\bibnamefont
  {Wallden}},\ }\bibfield  {booktitle} {\emph {\bibinfo {booktitle}
  {{Proceedings, 15th Conference on Recent Developments in Gravity (NEB 15):
  Chania, Crete, Greece, June 20-23, 2012}}},\ }\href {\doibase
  10.1088/1742-6596/453/1/012023} {\bibfield  {journal} {\bibinfo  {journal}
  {J. Phys. Conf. Ser.}\ }\textbf {\bibinfo {volume} {453}},\ \bibinfo {pages}
  {012023} (\bibinfo {year} {2013})}\BibitemShut {NoStop}%
\bibitem [{\citenamefont {Henson}(2009)}]{Henson:2006kf}%
  \BibitemOpen
  \bibfield  {author} {\bibinfo {author} {\bibfnamefont {J.}~\bibnamefont
  {Henson}},\ }in\ \href@noop {} {\emph {\bibinfo {booktitle} {Approaches to
  Quantum Gravity: Toward a New Understanding of Space, Time and Matter}}},\
  \bibinfo {editor} {edited by\ \bibinfo {editor} {\bibfnamefont
  {D.}~\bibnamefont {Oriti}}}\ (\bibinfo  {publisher} {Cambridge University
  Press},\ \bibinfo {year} {2009})\ pp.\ \bibinfo {pages} {393--413},\ \Eprint
  {http://arxiv.org/abs/gr-qc/0601121} {arXiv:gr-qc/0601121 [gr-qc]}
  \BibitemShut {NoStop}%
\bibitem [{\citenamefont {Amelino-Camelia}(2013)}]{AmelinoCamelia:2008qg}%
  \BibitemOpen
  \bibfield  {author} {\bibinfo {author} {\bibfnamefont {G.}~\bibnamefont
  {Amelino-Camelia}},\ }\href {\doibase 10.12942/lrr-2013-5} {\bibfield
  {journal} {\bibinfo  {journal} {Living Rev.Rel.}\ }\textbf {\bibinfo {volume}
  {16}},\ \bibinfo {pages} {5} (\bibinfo {year} {2013})},\ \Eprint
  {http://arxiv.org/abs/0806.0339} {arXiv:0806.0339 [gr-qc]} \BibitemShut
  {NoStop}%
\bibitem [{\citenamefont {Amelino-Camelia}\ \emph
  {et~al.}(2011{\natexlab{a}})\citenamefont {Amelino-Camelia}, \citenamefont
  {Freidel}, \citenamefont {Kowalski-Glikman},\ and\ \citenamefont
  {Smolin}}]{AmelinoCamelia:2011bm}%
  \BibitemOpen
  \bibfield  {author} {\bibinfo {author} {\bibfnamefont {G.}~\bibnamefont
  {Amelino-Camelia}}, \bibinfo {author} {\bibfnamefont {L.}~\bibnamefont
  {Freidel}}, \bibinfo {author} {\bibfnamefont {J.}~\bibnamefont
  {Kowalski-Glikman}}, \ and\ \bibinfo {author} {\bibfnamefont
  {L.}~\bibnamefont {Smolin}},\ }\href {\doibase 10.1103/PhysRevD.84.084010}
  {\bibfield  {journal} {\bibinfo  {journal} {Phys. Rev.}\ }\textbf {\bibinfo
  {volume} {D84}},\ \bibinfo {pages} {084010} (\bibinfo {year}
  {2011}{\natexlab{a}})},\ \Eprint {http://arxiv.org/abs/1101.0931}
  {arXiv:1101.0931 [hep-th]} \BibitemShut {NoStop}%
\bibitem [{\citenamefont {Born}(1938)}]{Born:1938}%
  \BibitemOpen
  \bibfield  {author} {\bibinfo {author} {\bibfnamefont {M.}~\bibnamefont
  {Born}},\ }\href {\doibase 10.1098/rspa.1938.0060} {\bibfield  {journal}
  {\bibinfo  {journal} {Proceedings of the Royal Society of London. Series A.
  Mathematical and Physical Sciences}\ }\textbf {\bibinfo {volume} {165}},\
  \bibinfo {pages} {291} (\bibinfo {year} {1938})}\BibitemShut {NoStop}%
\bibitem [{\citenamefont {Snyder}(1947)}]{Snyder:1946qz}%
  \BibitemOpen
  \bibfield  {author} {\bibinfo {author} {\bibfnamefont {H.~S.}\ \bibnamefont
  {Snyder}},\ }\href {\doibase 10.1103/PhysRev.71.38} {\bibfield  {journal}
  {\bibinfo  {journal} {Phys. Rev.}\ }\textbf {\bibinfo {volume} {71}},\
  \bibinfo {pages} {38} (\bibinfo {year} {1947})}\BibitemShut {NoStop}%
\bibitem [{\citenamefont {Majid}(2000)}]{Majid:1999tc}%
  \BibitemOpen
  \bibfield  {author} {\bibinfo {author} {\bibfnamefont {S.}~\bibnamefont
  {Majid}},\ }\bibfield  {booktitle} {\emph {\bibinfo {booktitle} {{Towards
  quantum gravity. Proceedings, 35th International Winter School on theoretical
  physics, Polanica, Poland, February 2-11, 1999}}},\ }\href@noop {} {\bibfield
   {journal} {\bibinfo  {journal} {Lect. Notes Phys.}\ }\textbf {\bibinfo
  {volume} {541}},\ \bibinfo {pages} {227} (\bibinfo {year} {2000})},\ \bibinfo
  {note} {[,227(1999)]},\ \Eprint {http://arxiv.org/abs/hep-th/0006166}
  {arXiv:hep-th/0006166 [hep-th]} \BibitemShut {NoStop}%
\bibitem [{\citenamefont {Lukierski}\ \emph {et~al.}(1991)\citenamefont
  {Lukierski}, \citenamefont {Ruegg}, \citenamefont {Nowicki},\ and\
  \citenamefont {Tolstoi}}]{Lukierski:1991pn}%
  \BibitemOpen
  \bibfield  {author} {\bibinfo {author} {\bibfnamefont {J.}~\bibnamefont
  {Lukierski}}, \bibinfo {author} {\bibfnamefont {H.}~\bibnamefont {Ruegg}},
  \bibinfo {author} {\bibfnamefont {A.}~\bibnamefont {Nowicki}}, \ and\
  \bibinfo {author} {\bibfnamefont {V.~N.}\ \bibnamefont {Tolstoi}},\ }\href
  {\doibase 10.1016/0370-2693(91)90358-W} {\bibfield  {journal} {\bibinfo
  {journal} {Phys. Lett.}\ }\textbf {\bibinfo {volume} {B264}},\ \bibinfo
  {pages} {331} (\bibinfo {year} {1991})}\BibitemShut {NoStop}%
\bibitem [{\citenamefont {Majid}\ and\ \citenamefont
  {Ruegg}(1994)}]{Majid1994}%
  \BibitemOpen
  \bibfield  {author} {\bibinfo {author} {\bibfnamefont {S.}~\bibnamefont
  {Majid}}\ and\ \bibinfo {author} {\bibfnamefont {H.}~\bibnamefont {Ruegg}},\
  }\href {\doibase 10.1016/0370-2693(94)90699-8} {\bibfield  {journal}
  {\bibinfo  {journal} {Phys. Lett.}\ }\textbf {\bibinfo {volume} {B334}},\
  \bibinfo {pages} {348} (\bibinfo {year} {1994})},\ \Eprint
  {http://arxiv.org/abs/hep-th/9405107} {arXiv:hep-th/9405107 [hep-th]}
  \BibitemShut {NoStop}%
\bibitem [{\citenamefont {Kowalski-Glikman}(2002)}]{KowalskiGlikman:2002ft}%
  \BibitemOpen
  \bibfield  {author} {\bibinfo {author} {\bibfnamefont {J.}~\bibnamefont
  {Kowalski-Glikman}},\ }\href {\doibase 10.1016/S0370-2693(02)02762-4}
  {\bibfield  {journal} {\bibinfo  {journal} {Phys. Lett.}\ }\textbf {\bibinfo
  {volume} {B547}},\ \bibinfo {pages} {291} (\bibinfo {year} {2002})},\ \Eprint
  {http://arxiv.org/abs/hep-th/0207279} {arXiv:hep-th/0207279 [hep-th]}
  \BibitemShut {NoStop}%
\bibitem [{\citenamefont {Majid}(1995)}]{Majid:1995qg}%
  \BibitemOpen
  \bibfield  {author} {\bibinfo {author} {\bibfnamefont {S.}~\bibnamefont
  {Majid}},\ }\href@noop {} {\emph {\bibinfo {title} {Foundations of Quantum
  Group Theory}}}\ (\bibinfo  {publisher} {Cambridge University Press},\
  \bibinfo {year} {1995})\BibitemShut {NoStop}%
\bibitem [{\citenamefont {Kowalski-Glikman}\ and\ \citenamefont
  {Nowak}(2002)}]{KowalskiGlikman:2002we}%
  \BibitemOpen
  \bibfield  {author} {\bibinfo {author} {\bibfnamefont {J.}~\bibnamefont
  {Kowalski-Glikman}}\ and\ \bibinfo {author} {\bibfnamefont {S.}~\bibnamefont
  {Nowak}},\ }\href {\doibase 10.1016/S0370-2693(02)02063-4} {\bibfield
  {journal} {\bibinfo  {journal} {Phys. Lett.}\ }\textbf {\bibinfo {volume}
  {B539}},\ \bibinfo {pages} {126} (\bibinfo {year} {2002})},\ \Eprint
  {http://arxiv.org/abs/hep-th/0203040} {arXiv:hep-th/0203040 [hep-th]}
  \BibitemShut {NoStop}%
\bibitem [{\citenamefont {Carmona}\ \emph {et~al.}(2016)\citenamefont
  {Carmona}, \citenamefont {Cortes},\ and\ \citenamefont
  {Relancio}}]{Carmona2016}%
  \BibitemOpen
  \bibfield  {author} {\bibinfo {author} {\bibfnamefont {J.~M.}\ \bibnamefont
  {Carmona}}, \bibinfo {author} {\bibfnamefont {J.~L.}\ \bibnamefont {Cortes}},
  \ and\ \bibinfo {author} {\bibfnamefont {J.~J.}\ \bibnamefont {Relancio}},\
  }\href {\doibase 10.1103/PhysRevD.94.084008} {\bibfield  {journal} {\bibinfo
  {journal} {Phys. Rev.}\ }\textbf {\bibinfo {volume} {D94}},\ \bibinfo {pages}
  {084008} (\bibinfo {year} {2016})},\ \Eprint
  {http://arxiv.org/abs/1609.01347} {arXiv:1609.01347 [hep-th]} \BibitemShut
  {NoStop}%
\bibitem [{\citenamefont {Amelino-Camelia}\ \emph
  {et~al.}(2011{\natexlab{b}})\citenamefont {Amelino-Camelia}, \citenamefont
  {Freidel}, \citenamefont {Kowalski-Glikman},\ and\ \citenamefont
  {Smolin}}]{AmelinoCamelia:2011pe}%
  \BibitemOpen
  \bibfield  {author} {\bibinfo {author} {\bibfnamefont {G.}~\bibnamefont
  {Amelino-Camelia}}, \bibinfo {author} {\bibfnamefont {L.}~\bibnamefont
  {Freidel}}, \bibinfo {author} {\bibfnamefont {J.}~\bibnamefont
  {Kowalski-Glikman}}, \ and\ \bibinfo {author} {\bibfnamefont
  {L.}~\bibnamefont {Smolin}},\ }\href {\doibase 10.1142/S0218271811020743,
  10.1007/s10714-011-1212-8} {\bibfield  {journal} {\bibinfo  {journal}
  {Gen.Rel.Grav.}\ }\textbf {\bibinfo {volume} {43}},\ \bibinfo {pages} {2547}
  (\bibinfo {year} {2011}{\natexlab{b}})},\ \Eprint
  {http://arxiv.org/abs/1106.0313} {arXiv:1106.0313 [hep-th]} \BibitemShut
  {NoStop}%
\bibitem [{\citenamefont {Lobo}\ and\ \citenamefont
  {Palmisano}(2016)}]{Lobo:2016blj}%
  \BibitemOpen
  \bibfield  {author} {\bibinfo {author} {\bibfnamefont {I.~P.}\ \bibnamefont
  {Lobo}}\ and\ \bibinfo {author} {\bibfnamefont {G.}~\bibnamefont
  {Palmisano}},\ }\bibfield  {booktitle} {\emph {\bibinfo {booktitle}
  {{Proceedings, 9th Alexander Friedmann International Seminar on Gravitation
  and Cosmology and 3rd Satellite Symposium on the Casimir Effect: St.
  Petersburg, Russia, June 21-27, 2015}}},\ }\href {\doibase
  10.1142/S2010194516601265} {\bibfield  {journal} {\bibinfo  {journal} {Int.
  J. Mod. Phys. Conf. Ser.}\ }\textbf {\bibinfo {volume} {41}},\ \bibinfo
  {pages} {1660126} (\bibinfo {year} {2016})},\ \Eprint
  {http://arxiv.org/abs/1612.00326} {arXiv:1612.00326 [hep-th]} \BibitemShut
  {NoStop}%
\bibitem [{\citenamefont {Battisti}\ and\ \citenamefont
  {Meljanac}(2010)}]{Battisti:2010sr}%
  \BibitemOpen
  \bibfield  {author} {\bibinfo {author} {\bibfnamefont {M.~V.}\ \bibnamefont
  {Battisti}}\ and\ \bibinfo {author} {\bibfnamefont {S.}~\bibnamefont
  {Meljanac}},\ }\href {\doibase 10.1103/PhysRevD.82.024028} {\bibfield
  {journal} {\bibinfo  {journal} {Phys. Rev.}\ }\textbf {\bibinfo {volume}
  {D82}},\ \bibinfo {pages} {024028} (\bibinfo {year} {2010})},\ \Eprint
  {http://arxiv.org/abs/1003.2108} {arXiv:1003.2108 [hep-th]} \BibitemShut
  {NoStop}%
\bibitem [{\citenamefont {Meljanac}\ \emph {et~al.}(2009)\citenamefont
  {Meljanac}, \citenamefont {Meljanac}, \citenamefont {Samsarov},\ and\
  \citenamefont {Stojic}}]{Meljanac:2009ej}%
  \BibitemOpen
  \bibfield  {author} {\bibinfo {author} {\bibfnamefont {S.}~\bibnamefont
  {Meljanac}}, \bibinfo {author} {\bibfnamefont {D.}~\bibnamefont {Meljanac}},
  \bibinfo {author} {\bibfnamefont {A.}~\bibnamefont {Samsarov}}, \ and\
  \bibinfo {author} {\bibfnamefont {M.}~\bibnamefont {Stojic}},\ }\href@noop {}
  {\  (\bibinfo {year} {2009})},\ \Eprint {http://arxiv.org/abs/0909.1706}
  {arXiv:0909.1706 [math-ph]} \BibitemShut {NoStop}%
\bibitem [{\citenamefont {Chern}\ \emph {et~al.}(1999)\citenamefont {Chern},
  \citenamefont {Chen},\ and\ \citenamefont {Lam}}]{Chern:1999jn}%
  \BibitemOpen
  \bibfield  {author} {\bibinfo {author} {\bibfnamefont {S.~S.}\ \bibnamefont
  {Chern}}, \bibinfo {author} {\bibfnamefont {W.~H.}\ \bibnamefont {Chen}}, \
  and\ \bibinfo {author} {\bibfnamefont {K.~S.}\ \bibnamefont {Lam}},\
  }\href@noop {} {\emph {\bibinfo {title} {{Lectures on differential
  geometry}}}}\ (\bibinfo {year} {1999})\ \bibinfo {note} {, see Eqs. (1.30)
  and (1.31) of Chapter 6}\BibitemShut {NoStop}%
\bibitem [{\citenamefont {Gubitosi}\ and\ \citenamefont
  {Mercati}(2013)}]{Gubitosi:2013rna}%
  \BibitemOpen
  \bibfield  {author} {\bibinfo {author} {\bibfnamefont {G.}~\bibnamefont
  {Gubitosi}}\ and\ \bibinfo {author} {\bibfnamefont {F.}~\bibnamefont
  {Mercati}},\ }\href {\doibase 10.1088/0264-9381/30/14/145002} {\bibfield
  {journal} {\bibinfo  {journal} {Class. Quant. Grav.}\ }\textbf {\bibinfo
  {volume} {30}},\ \bibinfo {pages} {145002} (\bibinfo {year} {2013})},\
  \Eprint {http://arxiv.org/abs/1106.5710} {arXiv:1106.5710 [gr-qc]}
  \BibitemShut {NoStop}%
\bibitem [{\citenamefont {Banburski}\ and\ \citenamefont
  {Freidel}(2014)}]{Banburski:2013jfa}%
  \BibitemOpen
  \bibfield  {author} {\bibinfo {author} {\bibfnamefont {A.}~\bibnamefont
  {Banburski}}\ and\ \bibinfo {author} {\bibfnamefont {L.}~\bibnamefont
  {Freidel}},\ }\href {\doibase 10.1103/PhysRevD.90.076010} {\bibfield
  {journal} {\bibinfo  {journal} {Phys. Rev.}\ }\textbf {\bibinfo {volume}
  {D90}},\ \bibinfo {pages} {076010} (\bibinfo {year} {2014})},\ \Eprint
  {http://arxiv.org/abs/1308.0300} {arXiv:1308.0300 [gr-qc]} \BibitemShut
  {NoStop}%
\bibitem [{\citenamefont {Amelino-Camelia}\ \emph {et~al.}(2016)\citenamefont
  {Amelino-Camelia}, \citenamefont {Gubitosi},\ and\ \citenamefont
  {Palmisano}}]{Amelino-Camelia:2013sba}%
  \BibitemOpen
  \bibfield  {author} {\bibinfo {author} {\bibfnamefont {G.}~\bibnamefont
  {Amelino-Camelia}}, \bibinfo {author} {\bibfnamefont {G.}~\bibnamefont
  {Gubitosi}}, \ and\ \bibinfo {author} {\bibfnamefont {G.}~\bibnamefont
  {Palmisano}},\ }\href {\doibase 10.1142/S0218271816500279} {\bibfield
  {journal} {\bibinfo  {journal} {Int. J. Mod. Phys.}\ }\textbf {\bibinfo
  {volume} {D25}},\ \bibinfo {pages} {1650027} (\bibinfo {year} {2016})},\
  \Eprint {http://arxiv.org/abs/1307.7988} {arXiv:1307.7988 [gr-qc]}
  \BibitemShut {NoStop}%
\end{thebibliography}
\end{document}